\documentclass[showpacs,preprintnumbers,%
amsmath,amssymb,aps,floatfix]{revtex4}

\usepackage{graphicx}
\usepackage{dcolumn}
\usepackage{bm}

 \newcommand{\vek}    [1] {\textrm{\textbf{{#1}}}} 
 \newcommand{\vekk}   [1] {{\bm  #1} } 
 \newcommand{\op}     [1] {\hat{\textrm{#1}}}  

 \newcommand{\me}         {\textrm{e}}   
 \newcommand{\mi}         {\textrm{i}}   
 \newcommand{\md}         {\,\textrm{d}} 
 \newcommand{\mean}   [1] {\langle #1 \rangle}
 \newcommand{\ket}    [1] {| #1 \rangle}
 \newcommand{\bra}    [1] {\langle #1 |}
 \newcommand{\bk}     [2] {\langle #1 | #2 \rangle}
 \newcommand{\kb}     [2] {| #1 \rangle \langle #2 |}
 \newcommand{\bok}    [3] {\langle #1 | #2 | #3 \rangle}

\begin{document}

\title{Generalized gauge-invariant formulations of the strong-field 
approximation}  

\author{Yulian V. Vanne}
\author{Alejandro Saenz}%
\affiliation{%
AG Moderne Optik, Institut f\"ur Physik, Humboldt-Universit\"at
         zu Berlin, 
         Hausvogteiplatz 5-7, D\,--\,10\,117 Berlin, Germany}%

\date{\today}%

\begin{abstract}
The gauge problem in the so-called strong-field approximation (SFA) 
describing atomic or molecular systems exposed to intense laser 
fields is investigated. Introducing a generalized gauge and 
partitioning of the Hamiltonian it is demonstrated that the 
$S$-matrix expansion obtained in the SFA depends on both gauge 
and partitioning in such a way that two gauges always yield the 
same $S$-matrix expansion, if the partitioning is properly chosen. 
\end{abstract}


\maketitle

\section{\label{sec:Intro}Introduction} 

Gauge-invariance is one of the fundamental concepts of electrodynamics. As 
a consequence it is, e.\,g., possible to formulate the interaction of 
charged particles with electromagnetic fields in different gauges. Although 
the choice of the gauge clearly influences parameters like the scalar or 
the vector potential, all physical quantities (observables) are independent 
of the gauge, if a complete treatment is performed. On the other hand, 
an approximate treatment often leads to gauge-dependent predictions 
for physical observables. One prominent example is given by the so-called  
strong-field approximation (SFA) for describing atomic or molecular systems 
exposed to intense laser fields that is also known as Keldysh-Faisal-Reiss 
(KFR) theory \cite{sfa:keld65,sfa:fais73,sfa:reis80}. 
It is based on an (infinite) series expansion of the exact $S$ matrix 
describing the interaction of an atomic system with a laser pulse. Besides 
a phase factor, the $S$ matrix obtained from a calculation of the complete 
expansion should thus be independent on the chosen gauge, provided the 
series converge. Consequently, the corresponding observable transition 
probabilities (obtained from the squared absolute values of the $S$-matrix 
elements) should be gauge independent. 

This gauge-independence of physical observables is usually lost, if 
only a truncated series is considered. This is the case for the SFA 
which is defined as the first term of the $S$-matrix expansion. 
In a number of investigations it has been shown that 
transition probabilities or rates predicted by either the length- or the 
velocity-gauge formulation of the SFA differ easily by one or two orders 
of magnitude for experimentally relevant laser parameters~\cite{sfa:baue06}. 
Furthermore, it was shown that the velocity-gauge SFA rate 
does not converge to the tunneling limit for weak fields, if long-range 
Coulomb interactions are present~\cite{sfa:vann07a}.
Recently, it was also demonstrated that there are pronounced 
qualitative differences between the energy distributions of the electrons 
ejected from, e.\,g., the 2p state of hydrogen atoms exposed to intense 
laser fields, if they are calculated within the SFA and either the length 
or the velocity gauge \cite{sfa:baue05b}.
The recent extensions of the 
SFA to molecular systems in velocity gauge \cite{sfm:muth00}, length 
gauge \cite{sfm:kjel04}, or dressed length gauge \cite{sfm:milo06} 
indicate similar or even more pronounced gauge dependencies for molecules.  
These findings have intensified the discussions whether 
the formulation of the SFA in one of the two gauges may be ``superior'' 
to the other. One approach to answer this question is rather pragmatic 
and based on a direct comparison of the SFA predictions in both gauges to 
either exact (numerical) solutions of the full time-dependent Schr\"odinger 
equation \cite{sfa:baue05b,sfm:awas08} or to experimental 
results \cite{sfm:kjel05a,sfa:berg07}. 
Clearly, if there is no {\it a priori} reason that 
one of the two SFA formulations is superior to the other, the conclusions 
may vary depending on the considered atomic or molecular system (even its 
quantum state) as well as on laser-pulse parameters. 

A second line of argumentation in favor of one of the two gauges is based 
on the question of ``universality''. For example, it has been argued that 
an evident limitation of the length-gauge formulation is the fact that 
in this case the predicted observables depend only on the scalar potential 
and thus a one-dimensional parameter, while the full description of an 
electromagnetic field requires in principle more than one dimension 
\cite{sfa:reis07a}. Very recently, Faisal proposes  
a ``gauge-invariant'' intense-field $S$-matrix theory that yields 
equal transition probabilities in length or velocity gauge; independent 
of the order of expansion  \cite{sfa:fais07}. Consequently, Faisal claims 
that his theory overcomes the above-mentioned long-standing discrepancy 
between the SFA in the two gauges. According to the findings in 
\cite{sfa:fais07} the ``gauge-invariant'' and thus universal $S$-matrix 
theory appears to be equivalent to the traditional length-gauge formulation. 
Specifically, the first-order term reproduces exactly the Keldysh result 
\cite{sfa:keld65} which was obtained in length gauge. In view of the 
popularity of the SFA for describing atomic and molecular ionization in 
intense laser fields (see, e.\,g., \cite{sfa:beck05} or \cite{sfa:milo06} 
and references therein), this is an important result. 

Besides the evident appeal of a universal $S$-matrix formulation which would 
provide an end to the long-lasting debates on the choice of the appropriate 
gauge (for a very recent example, see \cite{sfa:reis08,sfa:berg08}), 
it appears, however, quite surprising that such a formulation should 
exist. In fact, Faisal derives in \cite{sfa:fais07a} an alternative 
$S$-matrix expansion that reproduces in length and velocity gauge 
the traditional velocity-gauge result. This alternative series expansion 
is obtained by a different partioning of the Hamiltonian. The particular  
results obtained in \cite{sfa:fais07} and \cite{sfa:fais07a} immediately 
lead to the question of their generality. In the present work it is shown 
that it is possible to achieve an $S$-matrix expansion in agreement to any  
traditional SFA formulation in either length, velocity, or radiation 
gauge for an arbitrary choice of the gauge, if the Hamiltonian is 
correspondingly partitioned. In fact, introducing a generalized gauge 
transformation that includes the mentioned particular gauges as special 
cases, an in principle infinite set of different $S$-matrix expansions 
can be formulated. All of them can be shown to be achievable within any 
gauge as long as a proper partioning of the Hamiltonian is performed.  
Furthermore, the introduction of the generalized gauge allows to 
clearly demonstrate, how the choice of the gauge and the partitioning 
of the Hamiltonian describing the atomic or molecular 
system exposed to a laser field are connected with each other. This 
provides a much deeper insight in the gauge-problem of SFA that in 
fact turns out to be more properly described as an expansion problem. 

In order to provide a clear definition of terms and notations, the 
following section gives a brief overview over the gauges relevant to 
the present work, including brief discussions on (local) gauge invariance,  
form invariance, and the dipole approximation. Most importantly, a 
generalized gauge is introduced in Sec.\,\ref{sec:GenGauge}. 
Sec.\,\ref{sec:F:FE} discusses the wavefunction of a free-electron 
in an electromagnetic field and the influence of the chosen gauge on it. 
Equipped with these prerequisites, Sec.\,\ref{sec:Smatrix} discusses the 
$S$-matrix theory in different gauges. It represents thus the main 
results of this work in which it is demonstrated how various results 
($S$-matrix expansions) can be obtained using different combinations 
of gauge and partitioning.

\section{Gauges}

In the validity regime of the SFA the laser intensities are so high  
that the photon density is also very high. Under 
these circumstances the number of photons
can be treated as a continuous variable and the field can be described 
classically by using Maxwell's equations. Therefore,
in the following a \textit{semi-classical} theory will be used in which 
the radiation field is treated classically, but
the atomic or molecular system is described using quantum mechanics. 
The influence of the quantum system on the external field is also neglected.

\subsection{Local gauge invariance}%

A classical electromagnetic field is described by electric and magnetic
field vectors, $\vek{F}(\vek{r},t)$ and $\vek{B}(\vek{r},t)$ or, 
alternatively, by the scalar and vector potentials $\Phi(\vek{r},t)$ and 
$\vek{A}(\vek{r},t)$. Consider a system consisting of
an electron in a electrostatic potential $U(\vek{r})$ created by a nucleus 
(or some nuclei) which interacts 
with an external electromagnetic field. In semi-classical theory, the 
evolution of the system is governed by 
the time-dependent Schr\"odinger equation that satisfies 
\textit{local gauge invariance} and is given in the coordinate representation
with the minimal-coupling Hamiltonian by
\begin{equation}
  i \frac{\partial }{\partial t}\Psi_{\chi}(\vek{r},t) = 
  \left\{  \frac{1}{2} \Bigl[ \vek{p}_{\rm c} 
  + \vek{A}_{\chi}(\vek{r},t) \Bigr]^2 - \Phi_{\chi}(\vek{r},t) + U(\vek{r})
  \right\} \Psi_{\chi}(\vek{r},t)
\label{F:GITDSE}
\end{equation}
where the subscript denotes the used gauge $\chi$ and the operator of 
canonical momentum is given independently of the gauge as 
$\vek{p}_{\rm c} \equiv - i\nabla$; a consequence of the definition of 
the minimal-coupling Hamiltonian. The word \textit{invariance} means
that if the wavefunction $\Psi_{\chi}$ and both potentials 
$\vek{A}_{\chi}$ and $\Phi_{\chi}$ are simultaneously transformed 
into a new gauge $\chi'$ using the transformation recipes   
\begin{align}
\Psi_{\chi'}(\vek{r},t)   &= \Psi_{\chi}(\vek{r},t) 
                 e^{\left[ i T_{\chi \rightarrow \chi' }(\vek{r},t) \right]} &
\Psi_{\chi}(\vek{r},t)    &= \Psi_{\chi'}(\vek{r},t) 
                 e^{\left[ i T_{\chi' \rightarrow \chi }(\vek{r},t) \right]} \\
\vek{A}_{\chi'}(\vek{r},t)&= \vek{A}_{\chi}(\vek{r},t)  
                 -   \nabla T_{\chi \rightarrow \chi' }(\vek{r},t) &
\vek{A}_{\chi}(\vek{r},t) &= \vek{A}_{\chi'}(\vek{r},t) 
                 -   \nabla T_{\chi' \rightarrow \chi }(\vek{r},t) \\
\Phi_{\chi'}(\vek{r},t)   &= \Phi_{\chi}(\vek{r},t) 
       +  \frac{\partial}{\partial t} T_{\chi \rightarrow \chi' }(\vek{r},t) &  
\Phi_{\chi}(\vek{r},t)    &= \Phi_{\chi'}(\vek{r},t) 
       +  \frac{\partial}{\partial t} T_{\chi' \rightarrow \chi }(\vek{r},t)
       \quad ,
\label{F:GTrans}
\end{align}
equation~(\ref{F:GITDSE}) is transformed into itself,  
but with $\Psi_{\chi}\rightarrow \Psi_{\chi'}$,  
$\vek{A}_{\chi}\rightarrow \vek{A}_{\chi'}$, and 
$\Phi_{\chi}\rightarrow\Phi_{\chi'}$. The transformation functions 
$T_{\chi \rightarrow \chi' }(\vek{r},t)$ satisfy the relations
\begin{equation}
 T_{\chi \rightarrow \chi' } = - T_{\chi' \rightarrow \chi }, \quad 
 T_{\chi \rightarrow \chi' } + T_{\chi' \rightarrow \chi'' } 
 =  T_{\chi \rightarrow \chi''} \quad .
\label{}
\end{equation}

Of course, physical quantities as the
probability $P(\vek{r},t)$ or the electric and magnetic fields 
($\vek{F}(\vek{r},t)$ and $\vek{B}(\vek{r},t)$) are gauge 
independent, i.\,e.
\begin{align}
P(\vek{r},t)       &= |\Psi_{\chi}(\vek{r},t)|^2 =  
                                |\Psi_{\chi'}(\vek{r},t)|^2  \\
\vek{F}(\vek{r},t) &= -\nabla \Phi_{\chi}(\vek{r},t)  
        - \frac{\partial}{\partial t} \vek{A}_{\chi}(\vek{r},t)  
                    =       -\nabla \Phi_{\chi'}(\vek{r},t)  
        - \frac{\partial}{\partial t} \vek{A}_{\chi'}(\vek{r},t) \\
\vek{B}(\vek{r},t) &= \nabla  \times \vek{A}_{\chi}(\vek{r},t) 
                    = \nabla  \times \vek{A}_{\chi'}(\vek{r},t).
\label{F:Inv1}
\end{align}

\subsection{Form-invariant physical quantities}%

A form-invariant quantity is defined as a quantity whose corresponding 
operator
$G_{\chi} = G(\vek{A}_{\chi},\Phi_{\chi})$ is form invariant under a unitary 
transformation
$\op{T}_{\chi \rightarrow \chi' } =  
\exp \left[ i T_{\chi \rightarrow \chi' }(\vek{r},t) \right]$, i.\,e.,
\begin{equation}
G_{\chi'} \equiv \op{T}_{\chi \rightarrow \chi' } 
   G_{\chi} \op{T}_{\chi \rightarrow \chi' }^{\dagger} = 
   G(\vek{A}_{\chi'},\Phi_{\chi'})
\label{F:UnTr}
\end{equation}
The important difference between physical and non-physical quantities lies 
in the gauge invariance of the eigenvalues. The eigenvalues of a physical 
quantity (an observable) are identical in all gauges, whereas the 
eigenvalues of non-physical quantities may depend on the chosen gauge. It 
can be shown that the eigenvalues of a form-invariant quantity are 
gauge-invariant, i.\,e.\ a physical quantity must be form-invariant. 
Consider the operator $G_{\chi}$ with the eigenvalues 
$g_{n}$ and the corresponding eigenstates $|\chi,n\rangle$, 
\begin{equation}
G_{\chi}|\chi,n\rangle = g_{n} |\chi,n\rangle
\end{equation}
The operator $G_{\chi'}$ obtained by the unitary 
transformation~(\ref{F:UnTr}) has the eigenstates 
$|\chi',n\rangle = \op{T}_{\chi \rightarrow \chi' }|\chi,n\rangle$ and the 
same eigenvalues $g_{n}$
\begin{equation}
G_{\chi'}|\chi',n\rangle = (\op{T}_{\chi \rightarrow \chi' } 
                           G_{\chi} \op{T}_{\chi \rightarrow \chi' }^{\dagger})
                           ( \op{T}_{\chi \rightarrow \chi' } |\chi,n\rangle ) 
                         = \op{T}_{\chi \rightarrow \chi'} g_{n}|\chi,n\rangle  
                         = g_{n}|\chi',n\rangle 
\label{F:Gprime}
\end{equation}

Both $\vek{A}_{\chi}$ and $\Phi_{\chi}$ are not operators that correspond 
to physical quantities. The canonical momentum is also not form-invariant, 
\begin{equation}
\op{T}_{\chi \rightarrow \chi' } \vek{p}_{\rm c}  
     =   \op{T}_{\chi \rightarrow \chi'}^{\dagger} 
           \vek{p}_{\rm c} - \nabla T_{\chi \rightarrow \chi' }(\vek{r},t) \;, 
\label{}
\end{equation}
and, therefore, does not represent a physical measurable
quantity, whereas the mechanical momentum
\begin{equation}
 \vekk{\pi}_{\chi}  \equiv \vek{p}_{\rm c} + \vek{A}_{\chi}(\vek{r},t)
\label{F:mm}
\end{equation}
is form-invariant,
\begin{equation}
\op{T}_{\chi \rightarrow \chi' } \vekk{\pi}_{\chi} \op{T}_{\chi 
   \rightarrow \chi'}^{\dagger} 
   =   \vek{p}_{\rm c} - \nabla T_{\chi \rightarrow \chi' }(\vek{r},t)   
   + \vek{A}_{\chi} =  \vek{p}_{\rm c} + \vek{A}_{\chi'} = \vekk{\pi}_{\chi'},
\end{equation}
and represents an observable. The electrostatic potential
$U(\vek{r})$ also represents a physical quantity. The instantaneous energy 
operator of the system, $\mathcal{E}_{\chi}$,
\begin{equation}
\mathcal{E}_{\chi} \equiv  \frac{1}{2} \Bigl[ \vek{p}_{\rm c} 
      + \vek{A}_{\chi}(\vek{r},t) \Bigr]^2 + U(\vek{r})         
      \;=\;  \frac{\vekk{\pi}_{\chi}^2}{2} + U(\vek{r})
\label{F:ieo}
\end{equation}
represents a physical quantity, since it is a function of only physical 
quantities: $\vekk{\pi}_{\chi}$ and $U(\vek{r})$.

It is essential that both the total Hamiltonian of the system,
\begin{equation}
\op{H}_{\chi} \equiv   \frac{1}{2} \Bigl[ \vek{p}_{\rm c} 
   + \vek{A}_{\chi}(\vek{r},t) \Bigr]^2 - \Phi_{\chi}(\vek{r},t) + U(\vek{r})
   = \mathcal{E}_{\chi} - \Phi_{\chi}(\vek{r},t),
\label{F:th}
\end{equation}
and the field-free Hamiltonian,
\begin{equation}
\op{H}_0 \equiv \frac{\vek{p}_{\rm c}^2}{2} + U(\vek{r}),
\label{F:ffh}
\end{equation}
are not form-invariant, although they determine the evolution of a physical 
system.

\subsection{Radiation gauge}

The radiation gauge (labeled in the following by the subscript R) is 
convenient, if no sources are presented. It is defined by the relations
\begin{equation}
 \nabla\cdot\vek{A}_{\rm R} = 0, \quad \Phi_{\rm R} = 0 \quad .
\label{F:rg}
\end{equation}
Therefore, one has $\nabla \cdot (\vek{A}_{\rm R}\Psi) 
= \vek{A}_{\rm R} (\nabla\Psi) + (\nabla\cdot\vek{A}_{\rm R})\Psi 
= \vek{A}_{\rm R} (\nabla\Psi)$ and
the total Hamiltonian of the system possesses the form
\begin{equation}
\op{H}_{\rm R} =   \vek{p}_{\rm c}^2/2 + \vek{A}_{\rm R} \cdot \vek{p}_{\rm c}
  +  \vek{A}^2_{\rm R}/2    + U  \quad .
\label{F:thR}
\end{equation}

The vector potential in this gauge satisfies the wave equation and can be 
given by a superposition of monochromatic plane wave solutions. In the 
following, the vector function $\vek{A}(\vek{r},t)$ specifies the vector 
potential $\vek{A}_{\rm R}(\vek{r},t)$ in radiation gauge, so that
\begin{equation}
 \vek{F}(\vek{r},t) = - \frac{\partial}{\partial t} \vek{A}(\vek{r},t), \quad
 \vek{B}(\vek{r},t) = \nabla  \times \vek{A}(\vek{r},t).
\end{equation}

\subsection{Dipole approximation}

If the wavelength of the considered radiation is sufficiently long, the 
spatial variation of the radiation field across the system can be neglected. 
Assuming $\vek{A}({\bf r},t) \approx \vek{A}(t)$ one obtains
\begin{equation}
 \vek{F} = \vek{F}(t) = - \frac{\md \vek{A}(t)}{\md t}, \qquad \vek{B}  = 0.
\label{F:DA}
\end{equation}
There exist two gauges, length and velocity gauge, which are extensively 
used in the context of the dipole approximation. It is, however, 
known that the conditions for a breakdown of the dipole approximation differ in 
different gauges. In order to explicitly highlight the terms which are 
neglected in the subsequent use of the dipole approximation, the general 
definitions of length and velocity gauges are given in the following 
subsections.

\subsection{Velocity gauge}

The velocity gauge (labeled in the following by the subscript V) is often 
used to remove the square of the vector potential from the total 
Hamiltonian. Note, the name {\it velocity gauge} 
is frequently used, if in fact the radiation gauge is meant. The reason 
is the equivalence of these two gauges in the weak-field limit in which 
the $A^2$ term can be neglected (with respect to the remaining terms). 
For strong fields this term is, however, not negligible and it is thus 
important to distinguish between these two gauges. 
The velocity gauge can in fact be obtained from the radiation gauge 
by means of the transformation function 
\begin{equation}
  T_{{\rm R} \rightarrow {\rm V}}(\vek{r},t) = \frac{1}{2} \int^{t} 
                     \vek{A}^2(\vek{r},t') \md t' \equiv \beta(\vek{r},t)
                     \quad .
\label{F:RtoV}
\end{equation}
The vector and scalar potentials are given in the velocity gauge by 
\begin{equation}
 \vek{A}_{\rm V} = \vek{A} - \nabla \beta, \qquad
\Phi_{\rm V} = \vek{A}^2/2
\label{}
\end{equation}
and the total Hamiltonian can be rewritten as
\begin{equation}
\op{H}_{\rm V} =   \vek{p}_{\rm c}^2/2 +\vek{A}_{\rm V} \cdot \vek{p}_{\rm c} 
                   + \mi \Delta \beta  + U 
\label{F:thV}
\end{equation}
where the identity $\nabla\cdot\vek{A}_{\rm V} = - \Delta \beta$ has been used.

Within the dipole approximation, the vector potential has the simpler form
$\vek{A}_{\rm V} = \vek{A}(t)$ and Eq.\,(\ref{F:thV}) reduces to
\begin{equation}
\op{H}_{\rm V} =   \vek{p}_{\rm c}^2/2 + U + \vek{A}(t) \cdot \vek{p}_{\rm c} 
      \quad . 
\label{F:thVdip}
\end{equation}

\subsection{Length gauge} 

Another useful and popular gauge is the length gauge (labeled in the 
following by the subscript L). It is obtained from the radiation gauge,  
if the transformation function
\begin{equation}
  T_{{\rm R} \rightarrow {\rm L}}(\vek{r},t) = \vek{A}(\vek{r},t) \cdot \vek{r} 
\label{F:RtoL}
\end{equation}
is applied. This leads to the length-gauge vector and scalar potentials 
\begin{equation}
 \vek{A}_{\rm L} = \vek{A} - \nabla (\vek{A} \cdot \vek{r} ) = 
     - i \vek{L} \times \vek{A}  , \quad
     \Phi_{\rm L} = - \vek{F} \cdot \vek{r} 
\label{}
\end{equation}
where $\vek{L} \equiv \vek{r} \times \vek{p}_{\rm c}$ is the canonical angular 
momentum. Within the dipole approximation, the vector potential in length gauge
is simply zero. This leads to a simple form of the total Hamiltonian that is 
given in length gauge by 
\begin{equation}
\op{H}_{\rm L} =   \vek{p}_{\rm c}^2/2 + U + \vek{F}(t) \cdot \vek{r}.
\label{F:thLdip}
\end{equation}
Note, that in this case the operators of the mechanical and canonical momentum
coincide, $\vekk{\pi}_{\rm L} = \vek{p}_{\rm c}$.

\subsection{\label{sec:GenGauge}Generalized gauge} 

It can be shown that both the length and velocity gauges are only particular
cases of a generalized gauge defined by an arbitrary set of parameters, 
$X = \{x_1, x_2\}$. This gauge, which will be referred to as $X$ gauge, 
is obtained via the transformation
\begin{equation}
  T_{{\rm R} \rightarrow X}(\vek{r},t) = x_1 \vek{A}(t) \cdot \vek{r} + 
                                         x_2 \beta(t) 
\label{F:Rtog}
\end{equation}
In the $X$ gauge the vector and scalar potentials are given as 
\begin{equation}
 \vek{A}_{X}(t) = (1-x_1) \vek{A}(t)  , 
 \quad 
 \Phi_{X} = - x_1 \vek{F}(t) \cdot \vek{r} + x_2 \vek{A}^2(t)/2 \quad .
\label{}
\end{equation}
This leads to the total Hamiltonian
\begin{equation}
  \op{H}_{X} =   \vek{p}_{\rm c}^2/2 + U 
             + (1-x_1) \vek{A}(t) \cdot \vek{p}_{\rm c} 
             + x_1\, \vek{F}(t) \cdot \vek{r}
             + \left[ (1-x_1)^2 - x_2 \right] \vek{A}^2(t)/2 .
\label{F:thX}
\end{equation}
In particular, $X=\{0,0\}$, $X=\{0,1\}$, and $X=\{1,0\}$ correspond 
to the total Hamiltonians in radiation gauge (\ref{F:thR}), 
velocity gauge (\ref{F:thVdip}), and length gauge (\ref{F:thLdip}), 
respectively.

\section{\label{sec:F:FE}Free electron in a laser field}

\subsection{Volkov wave function}

Consider now a free electron in the presence of a laser field described  
by $\vek{A}(t)$. In velocity gauge the TDSE reads
\begin{equation}
  i \frac{\partial }{\partial t}\Psi_{\rm V}(\vek{r},t) =  
   \Bigl[ \vek{p}_{\rm c}^2/2 + \vek{A}(t) \cdot \vek{p}_{\rm c} \Bigr]
   \Psi_{V}(\vek{r},t).
\label{F:TDSEf}
\end{equation}
The solution of Eq.\,(\ref{F:TDSEf}) is the (non-relativistic) 
\textit{Volkov wavefunction}
\begin{equation}
 \Psi_{{\rm V},\vek{k}}(\vek{r},t) =  \exp\Bigl[ \mi \vek{k}
 \cdot  \bigl\{ \vek{r} - \vekk{\alpha}(t) \bigr\} - \mi E_k t \Bigr], 
\quad E_k = k^2/2.
\label{F:VWV}
\end{equation}
where $\vekk{\alpha}(t) = \int^t \vek{A}(t') \md t'$ and $\delta$-function 
normalization is used. It is important to emphasize the physical
meaning of the vector $\vek{k}$ and the scalar $E_k$ that are 
usually referred to as electron momentum and electron energy,
respectively. In fact, $\vek{k}$ is the mean value of canonical momentum 
in velocity gauge, $\mean{\vek{p}_{\rm c}}_{\rm V} \equiv \vek{k}$. The mean 
value of mechanical momentum is on the other hand time-dependent and equal to 
$\vek{k} + \vek{A}(t)$. As was discussed above, this latter value is 
gauge-independent and will be denoted in the following by $\vekk{\pi}$; 
thus one has $\vekk{\pi} = \vekk{\pi}(\vek{k},t) =  \vek{k} + \vek{A}(t)$. 
To stress the independence on a gauge, the vector function $\vekk{\pi}$ has 
no subscript, in contrast to the operators $\vekk{\pi}_{\rm R}$, 
$\vekk{\pi}_{\rm V}$, or $\vekk{\pi}_{\rm L}$. 
The mean value $\mathcal{E}(\vek{k},t) = \vekk{\pi}^2(\vek{k},t)/2$ of the 
instantaneous energy operator (\ref{F:ieo}) is thus also time-dependent.

Using the identity
\begin{equation}
  \frac{1}{2} \int^t \mathcal{E}(\vek{k},t') \md t' = E_k t + 
   \vek{k} \cdot \vekk{\alpha}(t) + \beta(t)
\end{equation}
the Volkov wavefunction can be written as either 
\begin{equation}
 \Psi_{{\rm R},\vek{k}}(\vek{r},t) =  \exp\Bigl[ \mi \vek{k} \cdot  \vek{r} - 
                 \mi \int^t \mathcal{E}(\vek{k},t') \md t'  \Bigr]
\label{F:VWR}
\end{equation} 
or 
\begin{equation}
 \Psi_{{\rm L},\vek{k}}(\vek{r},t) =  
                     \exp\Bigl[ \mi \vekk{\pi}(\vek{k},t) \cdot  \vek{r} - 
                      \mi \int^t \mathcal{E}(\vek{k},t') \md t'  \Bigr]
\label{F:VWL}
\end{equation} 
in radiation and length gauge, respectively. 

In $X$ gauge one finds 
\begin{equation}
 \Psi_{X,\vek{k}}(\vek{r},t) =  \exp\Bigl[ \mi  x_1 \vek{A}(t) \cdot \vek{r} + 
                                \mi x_2 \beta(t)  \Bigr]
 \Psi_{{\rm R},\vek{k}}(\vek{r},t) = \me^{-\mi \Theta_X(t)} 
                                \me^{ \mi  \vek{k} \cdot  \vek{r} }
\label{F:VWX}
\end{equation} 
with  
\begin{equation}
  \Theta_X(t) =  E_k t + \vek{k} \cdot \vekk{\alpha}(t) - 
                 x_1 \vek{A}(t) \cdot \vek{r} -  (x_2-1) \beta(t) \quad .
\label{ThetaX}
\end{equation}

Note, in length gauge $\vek{k}$ can be seen purely as a parameter, since
also the mean value of the canonical momentum $\mean{\vek{p}_{\rm c}}_{\rm L}$ 
is equal to $\vekk{\pi}$. Nevertheless, the physical meaning of $\vek{k}$
becomes evident in the case of a linear-polarized monochromatic 
electromagnetic field.

\subsection{Linear-polarized monochromatic field}

For a vector potential given by 
$\vek{A}(t) = - \vek{A}_0 \sin \omega t$ the corresponding electric field 
is $\vek{F}(t) = \vek{F}_0 \cos \omega t$ with $\vek{F}_0 = \omega \vek{A}_0$. 
In this case the mean value of the mechanical momentum is equal to 
$\vekk{\pi}(\vek{k},t) =  \vek{k} - \vek{F}_0/\omega \sin \omega t$.
Therefore, $\vek{k}$ is the value of the cycle-averaged $\vekk{\pi}$.  
An electron with $\vek{k} = 0$ quivers in the field around a single 
point in space.

The mean value $\mathcal{E}(\vek{k},t)$ of the instantaneous energy operator 
(\ref{F:ieo}) is time-dependent and given by
\begin{equation}
 \mathcal{E}(\vek{k},t) =  \frac{1}{2} \left(\vek{k} - \frac{\vek{F}_0}{\omega}
     \sin \omega t \right)^2 =  E_k - \vek{k} \cdot \frac{\vek{F}_0}{\omega} 
                                 \sin \omega t - U_p \cos 2 \omega t + U_p .
\label{F:ieoVW}
\end{equation}
Therefore, its cycle-averaged value is equal to $E_k + U_p$. The ponderomotive 
energy $U_p = F_0^2/(4\omega^2)$ is thus the cycle-averaged instantaneous 
energy of a quivering electron ($\vek{k} = 0$).

\section{\label{sec:Smatrix}Formal $S$-matrix formulation of SFA}

The following formulation of the $S$-matrix theory describing atomic 
and molecular systems in intense laser fields considers the case of a 
one-electron system for the sake of simplicity. The generalization 
to an arbitrary number of electrons is, however, straightforward.  
As a starting point the TDSE formulated in $X$ gauge is considered, 
\begin{equation}
\left( \mi \frac{\partial}{\partial t} - 
                             \op{H}_{X}(t) \right) \ket{\Psi_{X}(t)} = 0.
\label{}
\end{equation}
The electromagnetic field is absent before and after the pulse, i.\,e.
\begin{equation}
  \vek{A}(t) = \vek{F}(t) = 0,\quad \op{H}_{X}(t) = 
                       \op{H}{}^0 \qquad \text{for  $t < t_i$ and $t > t_f$}
\label{At0}
\end{equation}
Here, $\op{H}{}^0$ is the field-free Hamiltonian with 
eigenvalues $E_{\alpha}$ and eigenvectors $\ket{\psi_{\alpha}}$, 
\begin{equation}
 \op{H}{}^0 =   \vek{p}_{\rm c}^2/2 + U, 
         \qquad \op{H}{}^0\ket{\psi_{\alpha}} = E_{\alpha} \ket{\psi_{\alpha}}.
\label{}
\end{equation}
(The index $\alpha$ denotes discrete as well as continuum states and 
is thus itself either discrete or continuous.)

To describe the action of the pulse on the system, complete and orthonormal 
initial- and final-state basis sets are introduced. The initial-state basis 
set is given by 
$\ket{\psi_\alpha(t_i)} = \me^{- \mi E_\alpha t_i} \ket{\psi_{\alpha}}$
where the phase factor is introduced for convenience. The final-state basis 
set is given by plane waves with momentum $\vek{k}$, again for convenience 
multiplied by a phase factor, and depends both on the adopted gauge and 
on $\vek{k}$,
\begin{equation}
  \ket{\Psi_{X,\vek{k}}(t_f)} = \me^{-\mi \Theta_X(t_f)} \ket{\vek{k}}. 
\label{}
\end{equation}
The phase (see Eqs.\,(\ref{ThetaX}) and (\ref{At0}))
\begin{equation}
 \Theta_X(t_f) =  E_k t_f + \vek{k} \cdot \vekk{\alpha}(t_f) 
               -  (x_2-1) \beta(t_f)
\label{}
\end{equation}
is $\vek{r}$ independent but depends on the used gauge, $\vek{k}$, and the 
pulse. Note, that the for reasons of convenience introduced phase factors 
add only constant phases in the transition amplitudes and do not alter 
transition probabilities.

The probability amplitude of a transition from an initial state 
$\ket{\psi_\alpha(t_i)}$ to a final state $\ket{\Psi_{X,\vek{k}}(t_f)}$ 
is given by
\begin{equation}
  S_{\vek{k} \alpha} = 
      \mi \bok{ \Psi_{X,\vek{k}}(t_f) }{\op{G}_{X}(t_f,t_i)}{\psi_{\alpha}(t_i)}
\label{F:Smat}
\end{equation}
where the propagator  $\op{G}_{X}(t,t')$ is associated with $\op{H}_{X}(t)$ by 
the inhomogeneous equation
\begin{equation}
\left( \mi \frac{\partial}{\partial t} - 
                      \op{H}_{X}(t) \right) \op{G}_{X}(t,t') = \delta( t - t' ).
\label{}
\end{equation}

To obtain a systematic expansion of the transition amplitudes of interest it is 
convenient to express the total propagator $\op{G}_{X}$ of the system in terms 
of a partial propagator, defined by a \textit{partitioning} of the total 
Hamiltonian. The choice of the partitioning is made in such a way that 
the partial propagator can be expressed analytically, i.\,e.\ the 
Schr\"odinger equation with the corresponding partial Hamiltonian is 
solvable. 

A first class of Hamiltonians that leads to analytical solutions is 
the one describing a free electron in the field.
As was discussed in Sec.~\ref{sec:F:FE}, such Hamiltonians are 
gauge-dependent and their solutions are given by Volkov states. The 
partitioning of $\op{H}{}_{X}$ using the free-electron Hamiltonian in 
$X$ gauge, $\op{H}{}^{\rm f}_{X}$, is given by
\begin{equation}
  \op{H}{}_{X} = \op{H}{}^{\rm f}_{X} + U
\label{}
\end{equation}
The corresponding propagator can be written analytically using 
the solutions $\ket{\Psi_{X,\vek{k}}(t)}$,
\begin{equation}
 \op{G}{}^{\rm f}_{X}(t,t') = - i  \theta(t-t') \sum_{\vek{k}} 
                            \kb{ \Psi_{X,\vek{k}}(t) }{ \Psi_{X,\vek{k}}(t') },
\label{F:GfX}
\end{equation}
where $\theta(x)$ is the step function. From Eq.~(\ref{F:GfX}) follows
\begin{equation}
 - i \bra{\Psi_{X,\vek{k}}(t_f)}  \op{G}{}^{\rm f}_{X}(t_f,t) =  
      \bra{\Psi_{X,\vek{k}}(t)} \quad\text{for $t<t_f$}.
\label{F:PXGf}
\end{equation}

Another Hamiltonian that can be used for the partitioning is the field-free 
Hamiltonian  $\op{H}{}^0$. It is, however, only a special case of the class 
of Hamiltonians, which will be referred to as generalized field-free 
Hamiltonians and will be considered in the following subsection.

\subsection{Generalized field-free Hamiltonian}

The generalized field-free Hamiltonian
\begin{equation}
 \op{H}{}^0_{\gamma} = \me^{\mi \gamma(\vek{r},t)} \op{H}{}^0  
       \me^{- \mi \gamma(\vek{r},t)} - 
       \frac{\partial\gamma(\vek{r},t)}{\partial t} 
                     = \op{H}{}^0 + \frac{\mi}{2} \Delta\gamma 
        -  (\nabla\gamma) \cdot \vek{p}_{\rm c} 
        + \frac{1}{2}\left(  \nabla\gamma
                      \right)^2
        - \frac{\partial\gamma}{\partial t}
\label{}
\end{equation}
is defined with the aid of an arbitrary function $\gamma(\vek{r},t)$ in 
such a way that it reduces for $\gamma=0$ to the field-free Hamiltonian 
$\op{H}{}^0$. Solutions of the TDSE with $\op{H}{}^0_{\gamma}$ can be 
expanded in terms of the solutions of the TDSE with $\op{H}{}^0$ as
\begin{equation}
      \ket{\Psi_{\gamma,\alpha}(t)} = \me^{\mi \gamma(\vek{r},t)} 
                                 \me^{-\mi E_{\alpha} t} \ket{\psi_{\alpha}}.
\label{}
\end{equation}

In general, the function $\gamma(\vek{r},t)$ can be chosen independent of 
the gauge that is used to formulate the TDSE.
Consider a particular choice of $\gamma(\vek{r},t)$ parameterized 
by a set of parameters, $\lambda = \{\lambda_1, \lambda_2\}$,
\begin{equation}
  \gamma_{\lambda}(\vek{r},t) = \lambda_1 \vek{A}(t)\cdot\vek{r} + 
                                \lambda_2 \beta(t) \quad .
\label{}
\end{equation}
The corresponding generalized field-free Hamiltonian (the subscript
$\lambda$ is adopted instead of $\gamma_{\lambda}$ for the sake of 
notational simplicity) is then given by
\begin{equation}
     \op{H}{}^0_{\lambda} = \op{H}{}^0  
                          - \lambda_1\, \vek{A}(t) \cdot \vek{p}_{\rm c}  
                          + \lambda_1\, \vek{F}(t)\cdot\vek{r}
                          + ( \lambda_1^2 - \lambda_2 ) \vek{A}^2(t)/2.
\label{}
\end{equation}
Note, for all choices of $\lambda$ the Hamiltonian $\op{H}{}^0_{\lambda}$ 
gives an equivalent description of the evolution before and after the 
pulse, since for those times both $\vek{A}(t)$ and $\vek{F}(t)$ are equal 
to zero. Different choices of $\lambda$ yield, however, different partial 
propagators $\op{H}{}^0_{\lambda}$ during the pulse that can be written 
analytically as
\begin{equation}
 \op{G}{}^0_{\lambda}(t,t') = - i  \theta(t-t') \sum_{\alpha} 
                  \kb{ \Psi_{\lambda,\alpha}(t) }{ \Psi_{\lambda,\alpha}(t') }.
\label{F:G0l}
\end{equation}
In order to express  the total propagator $\op{G}{}_{X}$ in terms of 
$\op{G}{}^0_{\lambda}$, the total Hamiltonian is partitioned in two parts, 
\begin{equation}
     \op{H}_{X} = \op{H}{}^0_{\lambda} + V^0_{X,\lambda},
\label{}
\end{equation}
where the interaction operator $V^0_{X,\lambda}$ is given by
\begin{equation}
\begin{split}
 V^0_{X, \lambda}(t)  &= (1-x_1 + \lambda_1) \vek{A}(t) \cdot \vek{p}_{\rm c} 
      + (x_1 -  \lambda_1)\,  \vek{F}(t)\cdot\vek{r} \\
     & \qquad 
     + \left[ (1-x_1)^2 - x_2 - \lambda_1^2 + \lambda_2 \right] \vek{A}^2(t)/2
\label{}
\end{split}
\end{equation}

It is worth reminding that both sets of parameters, $X=\{x_1,x_2\}$ and 
$\lambda=\{\lambda_1, \lambda_2\}$, are independent of each other. 
Therefore, the same interaction operator can be obtained for different 
$X$ gauges, if the $\lambda$ parameters are appropriately chosen. 
It can be shown, for example, that  
\begin{align}
  V^0_{{\rm R}, \{-\!1,0 \}} &=  V^0_{{\rm V}, \{-1,1 \}} 
            =  V^0_{{\rm L}, \{0,0 \}} = \vek{F}(t)\cdot\vek{r} \\
  V^0_{{\rm R}, \{0,\!-\!1 \}} &=  V^0_{{\rm V}, \{ 0,0 \}} 
            =  V^0_{{\rm L}, \{1,1 \}} = \vek{A}(t)\cdot\vek{p}_{\rm c} \\
  V^0_{{\rm R}, \{ 0,0 \}\phantom{1}} &=  V^0_{{\rm V}, \{0,1 \}} =  
  V^0_{{\rm L}, \{1,2 \}} = \vek{A}(t)\cdot\vek{p}_{\rm c} + \vek{A}^2(t)/2
\label{}
\end{align}

Since $\ket{\Psi_{\lambda,\alpha}(t_i)} = \ket{\psi_\alpha(t_i)}$,  
Eq.~(\ref{F:G0l}) yields 
\begin{equation}
  i \op{G}{}^0_{\lambda}(t,t_i) \ket{\psi_\alpha(t_i)} =  
                   \ket{\Psi_{\lambda,\alpha}(t)}, \quad\text{for $t>t_i$}.
\label{F:G0Pl}
\end{equation}

\subsection{Matrix elements}

It will now be shown that most of the matrix elements of interest depend 
at most on the two parameters $v=\{v_1,v_2\}$ with 
$v_1 = 1 + \lambda_1 - x_1$ and $v_2 = \lambda_2 - x_2$.

Indeed, one finds for different matrix elements the relations 
\begin{equation}
 \bk{ \Psi_{X,\vek{k}}(t) }{ \Psi_{\lambda,\alpha}(t)} = 
        \bok{ \vek{k} }{\me^{\mi \Omega_{v\vek{k}\alpha}(t)}}{ \psi_{\alpha} },
\label{F:Xl}
\end{equation}
\begin{equation}
 \bok{ \Psi_{X,\vek{k}}(t) }{U}{ \Psi_{\lambda,\alpha}(t)} = 
        \bok{ \vek{k} }{\me^{\mi \Omega_{v\vek{k}\alpha}(t)}U}{ \psi_{\alpha} },
\label{F:XUl}
\end{equation}
\begin{equation}
 \bok{\Psi_{\lambda,\alpha'}(t)}{V^0_{X,\lambda}(t)}{\Psi_{\lambda,\alpha}(t)}= 
               \me^{\mi (E_{\alpha'} - E_{\alpha})t} 
               \bok{ \psi_{\alpha'} }{ \bar{V}^0_{v}(t) }{  \psi_{\alpha} },
\label{F:lVl}
\end{equation}
and
\begin{equation}
 \bok{ \Psi_{X,\vek{k}}(t) }{V^0_{X,\lambda}(t)}{ \Psi_{\lambda,\alpha}(t)} = 
          \bok{ \vek{k} }{\me^{\mi \Omega_{v\vek{k}\alpha}(t)} 
          \bar{V}^0_{v}(t)}{ \psi_{\alpha} }
\label{F:XVl}
\end{equation}
where
\begin{equation}
\begin{split}
\bar{V}^0_{v}(t)  &= V^0_{X,\lambda}(t) + \lambda_1 (1-x_1 + \lambda_1) 
                    \vek{A}^2(t)  \\
     &= v_1  \vek{A}(t) \cdot \vek{p}_{\rm c} 
      + ( 1 - v_1)  \vek{F}(t)\cdot\vek{r} + (v_1^2 + v_2) \vek{A}^2(t)/2 
\end{split}
\label{}
\end{equation}
and 
\begin{equation}
\begin{split}
  \Omega_{v\vek{k}\alpha}(t)  &= \Theta_X(t) + \gamma_{\lambda}(\vek{r},t) 
              - E_{\alpha} t \\
             &= (E_k - E_{\alpha}) t + \vek{k}\cdot\vekk{\alpha}(t) 
              + (v_1-1) \vek{A}(t)\cdot\vek{r} + (v_2+1) \beta(t)
\end{split}
\label{}
\end{equation}

Finally, the matrix element 
\begin{equation}
 \bok{ \Psi_{X,\vek{k}'}(t) }{U}{ \Psi_{X,\vek{k}}(t)} = 
    \me^{\mi (E_{k'} - E_k) t + \mi (\vek{k}'-\vek{k})\cdot \vekk{\alpha}(t)}
    \bok{ \vek{k}' }{U}{ \vek{k} }
\label{}
\end{equation}
is independent both on gauge and partitioning. 

As a consequence of these properties of the matrix elements the 
transition amplitude depends only on $v$, as is shown below.

\subsection{S-matrix series}

The operator $\op{G}_{X}(t,t')$ can be expanded either in terms of the 
operator $\op{G}{}^0_{\lambda}(t,t')$,
\begin{equation}
\op{G}_{X}(t,t') = \op{G}{}^0_{\lambda}(t,t') 
                 + \int \!\!\!\md t_1  \op{G}_{X}(t,t_1) 
                       V^0_{X,\lambda}(t_1) \op{G}{}^0_{\lambda}(t_1,t'),
\label{F:Sms1}
\end{equation}
or in terms of the operator $\op{G}{}^{\rm f}_{X}(t,t')$,
\begin{equation}
\op{G}_{X}(t,t') = \op{G}{}^{\rm f}_{X}(t,t') 
        + \int \!\!\!\md t_1  \op{G}{}^{\rm f}_{X}(t,t_1) U \op{G}_{X}(t_1,t').
\label{F:Sms2}
\end{equation}

Substitution of Eq.~(\ref{F:Sms2}) in (\ref{F:Sms1}) yields   
\begin{equation}
\begin{split}
\op{G}_{X}(t,t') &= \op{G}{}^0_{\lambda}(t,t') +
\int \!\!\!\md t_1 \op{G}{}^{\rm f}_{X}(t,t_1)  
                        V^0_{X,\lambda}(t_1) \op{G}{}^0_{\lambda}(t_1,t') \\
 &+ \iint \!\!\!\md t_2 \md t_1 \op{G}{}^{\rm f}_{X}(t,t_2) 
       U \op{G}_{X}(t_2,t_1) V^0_{X,\lambda}(t_1) \op{G}{}^0_{\lambda}(t_1,t').
\end{split}
\label{F:Sms3}
\end{equation}
A further substitution of either (\ref{F:Sms1}) or (\ref{F:Sms2}) in  
Eq.~(\ref{F:Sms3}) results in a series expansion of $\op{G}_{X}(t,t')$. 
Inserting this expansion in Eq.~(\ref{F:Smat}) generates  
the $S$-matrix series for the transition amplitude between the initial state  
and the final state to any desired order,
\begin{equation}
   S_{\vek{k}\alpha} = \sum_{n=0}^{\infty} S_{\vek{k}\alpha}^{(n)}, 
\label{}
\end{equation}
with
\begin{align}
  S_{\vek{k}\alpha}^{(0)} &= \mi\bok{ \Psi_{X,\vek{k}}(t_f) }{
                      \op{G}{}^0_{\lambda}(t_f,t_i)}{ \psi_{\alpha}(t_i) }, \\
  S_{\vek{k}\alpha}^{(1)} &= \mi\int \!\!\!\md t_1 \bok{ 
                         \Psi_{X,\vek{k}}(t_f) }{\op{G}{}^{\rm f}_{X}(t_f,t_1)  
     V^0_{X,\lambda}(t_1) \op{G}{}^0_{\lambda}(t_1,t_i)}{\psi_{\alpha}(t_i)}. \\
\intertext{Depending on whether (\ref{F:Sms1}) or (\ref{F:Sms2}) is 
substituted in (\ref{F:Sms3}) one obtains either}
\label{F:S2a}
  S_{\vek{k}\alpha}^{(2)} &= \mi \iint \!\!\!\md t_1 \md t_2 
      \bok{ \Psi_{X,\vek{k}}(t_f) }{\op{G}{}^{\rm f}_{X}(t_f,t_2) U  
      \op{G}{}^0_{\lambda}(t_2,t_1) V^0_{X,\lambda}(t_1) 
      \op{G}{}^0_{\lambda}(t_1,t_i)}{\psi_{\alpha}(t_i)} \\
\intertext{or}
\label{F:S2b}
  S_{\vek{k}\alpha}^{(2)} &= \mi \iint \!\!\!\md t_1 \md t_2 
           \bok{ \Psi_{X,\vek{k}}(t_f) }{\op{G}{}^{\rm f}_{X}(t_f,t_2)  
           U  \op{G}{}^{\rm f}_{X}(t_2,t_1) V^0_{X,\lambda}(t_1) 
           \op{G}{}^0_{\lambda}(t_1,t_i)}{\psi_{\alpha}(t_i)}
\end{align}
and so on, where the integration is performed in the range $t_i$ to $t_f$.

From Eqs.~(\ref{F:G0Pl}) and (\ref{F:Xl}) follows
\begin{equation}
S_{\vek{k}\alpha}^{(0)} = 
          \bk{\Psi_{X,\vek{k}}(t_f)}{\Psi_{\lambda,\alpha}(t_f)} 
       =  \me^{\mi \Omega_{v\vek{k}\alpha}(t_f)} \tilde{\psi}_{\alpha}(\vek{k})
\label{}
\end{equation}
where $\tilde{\psi}_{\alpha}(\vek{k}) = \bk{\vek{k}}{\psi_{\alpha}}$ is 
the Fourier transform of $\psi_{\alpha}$.

From Eqs.~(\ref{F:G0Pl}) and (\ref{F:PXGf}) follows on the other hand 
\begin{equation}
S_{\vek{k}\alpha}^{(1)} = \mi\int_{t_i}^{t_f} \!\!\!\md t 
 \bok{ \Psi_{X,\vek{k}}(t) }{ V^0_{X,\lambda}(t) }{\Psi_{\lambda,\alpha}(t)}
\label{}
\end{equation}
which --- using the identity (\ref{F:XVl}) --- can be reduced to
\begin{equation}
S_{\vek{k}\alpha}^{(1)} = \mi\int_{t_i}^{t_f} \!\!\!\md t \bok{ \vek{k} }{ 
     \me^{\mi \Omega_{v\vek{k}\alpha}(t)} \bar{V}^0_{v}(t)}{  \psi_{\alpha} }.
\label{}
\end{equation}

In an analogous way, Eq.~(\ref{F:S2a}) can be transformed using 
(\ref{F:G0l}) as
\begin{equation}
\begin{split}
S_{\vek{k}\alpha}^{(2)} &= \mi \int_{t_i}^{t_f} \!\!\!\md t_2 
              \int_{t_i}^{t_f} \!\!\!\md t_1 \bok{ \Psi_{X,\vek{k}}(t_2) }{
           U  \op{G}{}^0_{\lambda}(t_2,t_1) V^0_{X,\lambda}(t_1)  }{
           \psi_{\lambda,\alpha}(t_1)} \\
    &=  \int_{t_i}^{t_f} \!\!\!\md t_2  \sum_{\alpha'} 
        \bok{ \Psi_{X,\vek{k}}(t_2) }{U}{\Psi_{\lambda,\alpha'}(t_2)} 
        \int_{t_i}^{t_2} \!\!\!\md t_1 \bok{\Psi_{\lambda,\alpha'}(t_1)}{ 
        V^0_{X,\lambda}(t_1)  }{\Psi_{\lambda,\alpha}(t_1)}\\
    &= \int_{t_i}^{t_f} \!\!\!\md t_2  \sum_{\alpha'} 
        \bok{ \vek{k} }{\me^{\mi \Omega_{v\vek{k}\alpha'}(t_2)}U}{ 
        \psi_{\alpha'} }
        \int_{t_i}^{t_2} \!\!\!\md t_1 
        \me^{\mi (E_{\alpha'} - E_{\alpha})t_1} \bok{ \psi_{\alpha'} }{ 
        \bar{V}^0_{v}(t_1) }{  \psi_{\alpha} }
\end{split}
\label{}
\end{equation}
or Eq.~(\ref{F:S2b}) can be transformed using (\ref{F:GfX}) as
\begin{equation}
\begin{split}
S_{\vek{k}\alpha}^{(2)} &= \mi \int_{t_i}^{t_f} \!\!\!\md t_2 
         \int_{t_i}^{t_f}\!\!\! \md t_1 \bok{ \Psi_{X,\vek{k}}(t_2) }{ 
         U \op{G}{}^{\rm f}_{X}(t_2,t_1) V^0_{X,\lambda}(t_1)  }{
         \psi_{\lambda,\alpha}(t_1)} \\
         &=  \int_{t_i}^{t_f} \!\!\!\md t_2  \sum_{\vek{k}'} 
             \bok{ \Psi_{X,\vek{k}}(t_2) }{U}{\Psi_{X,\vek{k}'}(t_2)} 
            \int_{t_i}^{t_2} \!\!\!\md t_1 
             \bok{\Psi_{X,\vek{k}'}(t_1)}{ V^0_{X,\lambda}(t_1)  }{
             \Psi_{\lambda,\alpha}(t_1)} \\
         &=  \int_{t_i}^{t_f} \!\!\!\md t_2  \sum_{\vek{k}'} 
             \me^{\mi (E_{k} - E_{k'}) t_2 + 
             \mi (\vek{k}-\vek{k}')\cdot \vekk{\alpha}(t_2)}
             \bok{ \vek{k} }{U}{ \vek{k}' }
             \int_{t_i}^{t_2} \!\!\!\md t_1 
             \bok{ \vek{k}' }{\me^{\mi \Omega_{v\vek{k}'\alpha}(t_1)} 
             \bar{V}^0_{v}(t_1) }{ \psi_{\alpha} } .
\end{split}
\label{}
\end{equation}
Continuing in an analogous manner, it can be shown that 
$S_{\vek{k}\alpha}^{(n)}$ for any order $n$ depends only on $v$.
Therefore, $S_{\vek{k}\alpha}$ itself depends only on $v$. In the next 
subsection some particular cases will be considered explicitly.

\subsection{Particular cases}

As a first example, consider the case studied in 
\cite{sfa:fais07}. It is obtained using $v=\{0,0\}$ where one has 
\begin{align}
\bar{V}^0_{v}(t)  &= \vek{F}(t)\cdot\vek{r} \\
\Omega_{v\vek{k}\alpha}(t) & = (E_k - E_{\alpha}) t + 
         \vek{k}\cdot\vekk{\alpha}(t)   - \vek{A}(t)\cdot\vek{r} + \beta(t) .
\label{}
\end{align}
This formulation is achieved using the following partitionings for 
different gauges:
\begin{equation}
 \lambda=\{-1,0\} \text{ in R-gauge},\quad
 \lambda=\{-1,1\} \text{ in V-gauge},\quad
 \lambda=\{ 0,0\} \text{ in L-gauge}.  
\label{}
\end{equation}
Since in L gauge the relation $\op{H}{}^0_{\lambda} = \op{H}{}^0$ holds, 
the gauge-invariant formulation with $v=\{0,0\}$ reproduces the 
traditional SFA in the L gauge.

However, in an analogous way, the traditional V-gauge SFA is obtained with 
$v=\{1,-1\}$ (cf.\ \cite{sfa:fais07a}), where
\begin{align}
\bar{V}^0_{v}(t)  &= \vek{A}(t)\cdot\vek{p}_{\rm c}  \\
\Omega_{v\vek{k}\alpha}(t) & = (E_k - E_{\alpha}) t + 
          \vek{k}\cdot\vekk{\alpha}(t) \; .
\label{}
\end{align}
It can be achieved using the following partitionings for different gauges:
\begin{equation}
 \lambda=\{0,-1\} \text{ in R-gauge},\quad
 \lambda=\{0,0\} \text{ in V-gauge},\quad
 \lambda=\{1,-1\} \text{ in L-gauge}.  
\label{}
\end{equation}

In a similar way, the traditional R-gauge SFA is obtained with 
$v=\{1,0\}$, where
\begin{align}
\bar{V}^0_{v}(t)  &= \vek{A}(t)\cdot\vek{p}_{\rm c}  + \vek{A}^2(t)/2 \\
\Omega_{v\vek{k}\alpha}(t) & = (E_k - E_{\alpha}) t 
    + \vek{k}\cdot\vekk{\alpha}(t) +  \beta(t) .
\label{}
\end{align}
It can be achieved using the following partitionings for different gauges:
\begin{equation}
 \lambda=\{0,0\} \text{ in R-gauge},\quad
 \lambda=\{0,1\} \text{ in V-gauge},\quad
 \lambda=\{1,0\} \text{ in L-gauge}.  
\label{}
\end{equation}
Clearly, every $S$-matrix expansion (SFA formulation) in one of the 
``conventional'' (length, velocity, or radiation) gauges can be 
obtained by adopting any of the possible gauges, if the partitioning 
of the Hamiltonian is chosen accordingly.

\section{\label{sec:conclusion}Conclusion}

In this work it is shown how the (infinite-order) $S$-matrix expansion 
describing atomic or molecular systems in intense laser fields  
depends on the choice of both the gauge {\it and} the partitioning of the 
Hamiltonian. For this purpose a generalized gauge as well as a generalized 
partioning scheme is introduced. They are defined by 4 independent parameters 
($x_1$ and $x_2$ for the gauge as well as $\lambda_1$ and $\lambda_2$ for 
the partitioning). However, the $S$-matrix expansion is then shown to depend 
only on 2 parameters, $v_1=1+\lambda_1 - x_1$ and $v_2=\lambda_2-x_2$. 
Clearly, every possible combination of the parameters defining the gauge and 
the partitioning that conserves the values $\lambda_1 - x_1$ and 
$\lambda_2-x_2$ leads to an identical $S$-matrix expansion (up to an 
overall phase factor that cancels when calculating physical observables).    

The present analysis thus shows that one has to be very careful not to consider 
solely the gauge, since a suitable choice of the partitioning may lead 
to an $S$-matrix expansion that is identical to the one obtained in some 
other gauge, as was also demonstrated in \cite{sfa:fais07} where the 
velocity-gauge expansion appeared to agree with what is usually known as 
length-gauge $S$-matrix expansion. In fact, it is much more appropriate 
to discuss different expansions than different gauges. This avoids for 
example the rather contradictory terminology of a ``gauge-invariant 
(first-order) KFR approximation in the velocity gauge'' used 
in \cite{sfa:fais07}.  

Furthermore, the present result demonstrates that there 
remains an in principle infinite set of $S$-matrix expansions (characterized 
by different values of $v_1, v_2$) that are only shown to provide the 
same transition probabilities in the limit of an infinite series 
expansion, if the latter converges. Truncated series like, e.\,g., the 
0th, 1st, or 2nd  order expansions will, however, in general disagree. 
Thus the question of a ``proper'' choice of the expansion in the case 
of truncation remains and can only be clarified by a comparison to 
either experiment or gauge-independent theory (like full solutions 
of the time-dependent Schr\"odinger equation). It should be emphasized, 
however, that there is no {\it a priori} reason to believe that one 
expansion is necessarily advantageous to the others.  

\section*{Acknowledgments}
The authors would like to thank H.~R.~Reiss for in-depth discussions.
This work was supported by the {\it Deutsche
Forschungsgemeinschaft}. AS is grateful
to the {\it Stifterverband f\"ur die Deutsche Wissenschaft} (Programme
{\it Forschungsdozenturen}) and the {\it Fonds der Chemischen Industrie}
for financial support.



\begin{thebibliography}{19}
\expandafter\ifx\csname natexlab\endcsname\relax\def\natexlab#1{#1}\fi
\expandafter\ifx\csname bibnamefont\endcsname\relax
  \def\bibnamefont#1{#1}\fi
\expandafter\ifx\csname bibfnamefont\endcsname\relax
  \def\bibfnamefont#1{#1}\fi
\expandafter\ifx\csname citenamefont\endcsname\relax
  \def\citenamefont#1{#1}\fi
\expandafter\ifx\csname url\endcsname\relax
  \def\url#1{\texttt{#1}}\fi
\expandafter\ifx\csname urlprefix\endcsname\relax\def\urlprefix{URL }\fi
\providecommand{\bibinfo}[2]{#2}
\providecommand{\eprint}[2][]{\url{#2}}

\bibitem[{\citenamefont{Keldysh}(1965)}]{sfa:keld65}
\bibinfo{author}{\bibfnamefont{L.~V.} \bibnamefont{Keldysh}},
  \bibinfo{journal}{Sov.\,Phys.\ JETP} \textbf{\bibinfo{volume}{20}},
  \bibinfo{pages}{1307} (\bibinfo{year}{1965}).

\bibitem[{\citenamefont{Faisal}(1973)}]{sfa:fais73}
\bibinfo{author}{\bibfnamefont{F.~H.~M.} \bibnamefont{Faisal}},
  \bibinfo{journal}{J.\,Phys.\ B: At.\,Mol.\,Phys.}
  \textbf{\bibinfo{volume}{6}}, \bibinfo{pages}{L89} (\bibinfo{year}{1973}).

\bibitem[{\citenamefont{Reiss}(1980)}]{sfa:reis80}
\bibinfo{author}{\bibfnamefont{H.~R.} \bibnamefont{Reiss}},
  \bibinfo{journal}{Phys.\,Rev.\ A} \textbf{\bibinfo{volume}{22}},
  \bibinfo{pages}{1786} (\bibinfo{year}{1980}).

\bibitem[{\citenamefont{Bauer}(2006)}]{sfa:baue06}
\bibinfo{author}{\bibfnamefont{J.}~\bibnamefont{Bauer}},
  \bibinfo{journal}{Phys.\,Rev.\ A} \textbf{\bibinfo{volume}{73}},
  \bibinfo{pages}{023421} (\bibinfo{year}{2006}).

\bibitem[{\citenamefont{Vanne and Saenz}(2007)}]{sfa:vann07a}
\bibinfo{author}{\bibfnamefont{Y.~V.} \bibnamefont{Vanne}} \bibnamefont{and}
  \bibinfo{author}{\bibfnamefont{A.}~\bibnamefont{Saenz}},
  \bibinfo{journal}{Phys.\,Rev.\ A} \textbf{\bibinfo{volume}{75}},
  \bibinfo{pages}{063403} (\bibinfo{year}{2007}).

\bibitem[{\citenamefont{Bauer et~al.}(2005)\citenamefont{Bauer,
  {Milo\v{s}evi{\'c}}, and Becker}}]{sfa:baue05b}
\bibinfo{author}{\bibfnamefont{D.}~\bibnamefont{Bauer}},
  \bibinfo{author}{\bibfnamefont{D.~B.} \bibnamefont{{Milo\v{s}evi{\'c}}}},
  \bibnamefont{and} \bibinfo{author}{\bibfnamefont{W.}~\bibnamefont{Becker}},
  \bibinfo{journal}{Phys.\,Rev.\ A} \textbf{\bibinfo{volume}{72}},
  \bibinfo{pages}{023415} (\bibinfo{year}{2005}).

\bibitem[{\citenamefont{Muth-B{\"o}hm et~al.}(2000)\citenamefont{Muth-B{\"o}hm,
  Becker, and Faisal}}]{sfm:muth00}
\bibinfo{author}{\bibfnamefont{J.}~\bibnamefont{Muth-B{\"o}hm}},
  \bibinfo{author}{\bibfnamefont{A.}~\bibnamefont{Becker}}, \bibnamefont{and}
  \bibinfo{author}{\bibfnamefont{F.~H.~M.} \bibnamefont{Faisal}},
  \bibinfo{journal}{Phys.\,Rev.\,Lett.} \textbf{\bibinfo{volume}{85}},
  \bibinfo{pages}{2280} (\bibinfo{year}{2000}).

\bibitem[{\citenamefont{Kjeldsen and Madsen}(2004)}]{sfm:kjel04}
\bibinfo{author}{\bibfnamefont{T.~K.} \bibnamefont{Kjeldsen}} \bibnamefont{and}
  \bibinfo{author}{\bibfnamefont{L.~B.} \bibnamefont{Madsen}},
  \bibinfo{journal}{J.\,Phys.\ B: At.\,Mol.\,Phys.}
  \textbf{\bibinfo{volume}{37}}, \bibinfo{pages}{2033} (\bibinfo{year}{2004}).

\bibitem[{\citenamefont{Milo{\v{s}}evi{\'c}}(2006)}]{sfm:milo06}
\bibinfo{author}{\bibfnamefont{D.~B.} \bibnamefont{Milo{\v{s}}evi{\'c}}},
  \bibinfo{journal}{Phys.\,Rev.\ A} \textbf{\bibinfo{volume}{74}},
  \bibinfo{pages}{063404} (\bibinfo{year}{2006}).

\bibitem[{\citenamefont{Awasthi et~al.}(2008)\citenamefont{Awasthi, Vanne,
  Saenz, Castro, and Decleva}}]{sfm:awas08}
\bibinfo{author}{\bibfnamefont{M.}~\bibnamefont{Awasthi}},
  \bibinfo{author}{\bibfnamefont{Y.~V.} \bibnamefont{Vanne}},
  \bibinfo{author}{\bibfnamefont{A.}~\bibnamefont{Saenz}},
  \bibinfo{author}{\bibfnamefont{A.}~\bibnamefont{Castro}}, \bibnamefont{and}
  \bibinfo{author}{\bibfnamefont{P.}~\bibnamefont{Decleva}},
  \bibinfo{journal}{Phys.\,Rev.\ A} \textbf{\bibinfo{volume}{77}},
  \bibinfo{pages}{063403} (\bibinfo{year}{2008}).

\bibitem[{\citenamefont{Kjeldsen and Madsen}(2005)}]{sfm:kjel05a}
\bibinfo{author}{\bibfnamefont{T.~K.} \bibnamefont{Kjeldsen}} \bibnamefont{and}
  \bibinfo{author}{\bibfnamefont{L.~B.} \bibnamefont{Madsen}},
  \bibinfo{journal}{Phys.\,Rev.\ A} \textbf{\bibinfo{volume}{71}},
  \bibinfo{pages}{023411} (\bibinfo{year}{2005}).

\bibitem[{\citenamefont{Bergues et~al.}(2007)\citenamefont{Bergues, Ansari,
  Hanstorp, and Kiyan}}]{sfa:berg07}
\bibinfo{author}{\bibfnamefont{B.}~\bibnamefont{Bergues}},
  \bibinfo{author}{\bibfnamefont{Z.}~\bibnamefont{Ansari}},
  \bibinfo{author}{\bibfnamefont{D.}~\bibnamefont{Hanstorp}}, \bibnamefont{and}
  \bibinfo{author}{\bibfnamefont{I.~Y.} \bibnamefont{Kiyan}},
  \bibinfo{journal}{Phys.\,Rev.\ A} \textbf{\bibinfo{volume}{75}},
  \bibinfo{pages}{063415} (\bibinfo{year}{2007}).

\bibitem[{\citenamefont{Reiss}(2007)}]{sfa:reis07a}
\bibinfo{author}{\bibfnamefont{H.~R.} \bibnamefont{Reiss}},
  \bibinfo{journal}{Phys.\,Rev.\ A} \textbf{\bibinfo{volume}{76}},
  \bibinfo{pages}{033404} (\bibinfo{year}{2007}).

\bibitem[{\citenamefont{Faisal}(2007{\natexlab{a}})}]{sfa:fais07}
\bibinfo{author}{\bibfnamefont{F.~H.~M.} \bibnamefont{Faisal}},
  \bibinfo{journal}{J.\,Phys.\ B: At.\,Mol.\,Phys.}
  \textbf{\bibinfo{volume}{40}}, \bibinfo{pages}{F145}
  (\bibinfo{year}{2007}{\natexlab{a}}).

\bibitem[{\citenamefont{Becker and Faisal}(2005)}]{sfa:beck05}
\bibinfo{author}{\bibfnamefont{A.}~\bibnamefont{Becker}} \bibnamefont{and}
  \bibinfo{author}{\bibfnamefont{F.~H.~M.} \bibnamefont{Faisal}},
  \bibinfo{journal}{J.\,Phys.\ B: At.\,Mol.\,Phys.}
  \textbf{\bibinfo{volume}{38}}, \bibinfo{pages}{R1} (\bibinfo{year}{2005}).

\bibitem[{\citenamefont{{Milo\v{s}evi\'{c}}
  et~al.}(2006)\citenamefont{{Milo\v{s}evi\'{c}}, Paulus, Bauer, and
  Becker}}]{sfa:milo06}
\bibinfo{author}{\bibfnamefont{D.~B.} \bibnamefont{{Milo\v{s}evi\'{c}}}},
  \bibinfo{author}{\bibfnamefont{G.~G.} \bibnamefont{Paulus}},
  \bibinfo{author}{\bibfnamefont{D.}~\bibnamefont{Bauer}}, \bibnamefont{and}
  \bibinfo{author}{\bibfnamefont{W.}~\bibnamefont{Becker}},
  \bibinfo{journal}{J.\,Phys.\ B: At.\,Mol.\,Phys.}
  \textbf{\bibinfo{volume}{39}}, \bibinfo{pages}{R203} (\bibinfo{year}{2006}).

\bibitem[{\citenamefont{Reiss}(2008)}]{sfa:reis08}
\bibinfo{author}{\bibfnamefont{H.~R.} \bibnamefont{Reiss}},
  \bibinfo{journal}{Phys.\,Rev.\ A} \textbf{\bibinfo{volume}{77}},
  \bibinfo{pages}{067401} (\bibinfo{year}{2008}).

\bibitem[{\citenamefont{Bergues et~al.}(2008)\citenamefont{Bergues, Ansari,
  Hanstorp, and Kiyan}}]{sfa:berg08}
\bibinfo{author}{\bibfnamefont{B.}~\bibnamefont{Bergues}},
  \bibinfo{author}{\bibfnamefont{Z.}~\bibnamefont{Ansari}},
  \bibinfo{author}{\bibfnamefont{D.}~\bibnamefont{Hanstorp}}, \bibnamefont{and}
  \bibinfo{author}{\bibfnamefont{I.~Y.} \bibnamefont{Kiyan}},
  \bibinfo{journal}{Phys.\,Rev.\ A} \textbf{\bibinfo{volume}{77}},
  \bibinfo{pages}{067402} (\bibinfo{year}{2008}).

\bibitem[{\citenamefont{Faisal}(2007{\natexlab{b}})}]{sfa:fais07a}
\bibinfo{author}{\bibfnamefont{F.~H.~M.} \bibnamefont{Faisal}},
  \bibinfo{journal}{Phys.\,Rev.\ A} \textbf{\bibinfo{volume}{75}},
  \bibinfo{pages}{063412} (\bibinfo{year}{2007}{\natexlab{b}}).

\end{thebibliography}

\bibliographystyle{apsrev}


\end{document}